# Modulation of galactic protons in the heliosphere during the unusual solar minimum of 2006 to 2009

M. S. Potgieter[1], E. E. Vos[1],
M. Boezio[2], N. De Simone[3], V. Di Felice[3] and V. Formato[2]

1. Centre for Space Research, North-West University, 2520 Potchefstroom, South Africa
2. INFN, Sezione di Trieste, Padriciano 99, I-34012 Trieste, Italy
3. INFN, Sezione di Roma Tor Vergata, Via della Ricerca Scientifica 1, I-00133 Rome, Italy

**Abstract** The last solar minimum activity period, and the consequent minimum modulation conditions for cosmic rays, was unusual. The highest levels of galactic protons were recorded at Earth in late 2009 in contrast to expectations. Proton spectra observed for 2006 to 2009 from the *PAMELA* cosmic ray detector on-board the *Resurs-DK1* satellite are presented together with the solutions of a comprehensive numerical model for the solar modulation of cosmic rays. The model is used to determine what mechanisms were mainly responsible for the modulation of protons during this period, and why the observed spectrum for 2009 was the highest ever recorded. From mid-2006 until December 2009 we find that the spectra became significantly softer because increasingly more low energy protons had reached Earth. To simulate this effect, the rigidity dependence of the diffusion coefficients had to decrease significantly below ~3 GeV. The modulation minimum period of 2009 can thus be described as relatively more 'diffusion dominated' than previous solar minima. However, we illustrate that drifts still had played a significant role but that the observable modulation effects were not as well correlated with the waviness of the heliospheric current sheet as before. Protons still experienced global gradient and curvature drifts as the heliospheric magnetic field had decreased significantly until the end of 2009, in contrast to the moderate decreases observed during previous minimum periods. We conclude that all modulation processes contributed to the observed increases in the proton spectra for this period, exhibiting an intriguing interplay of these major mechanisms.

**Keywords** cosmic rays; solar modulation; solar minimum activity; galactic protons

[*]Corresponding author: Address: Tel.: +27 182992406; fax: +27 182992421.
E-mail: Marius.Potgieter@nwu.ac.za. (M.S. Potgieter)

## 1. Introduction

The past solar minimum activity period and the consequent minimum modulation conditions for galactic cosmic rays (CRs) had been unusual. It was expected that the new activity cycle would begin early in 2008, assuming a 10.5 year periodicity. Instead, solar minimum modulation conditions had continued until the end of 2009, characterized by a much weaker heliospheric magnetic field (HMF) compared to previous cycles. The tilt angle of the wavy heliospheric current sheet (HCS), on the other hand, had not decreased as rapidly as the magnitude of the HMF at Earth during this period, but eventually also reached a minimum value at the end of 2009. Mewaldt *et al.* (2010) and Ahluwalia and Ygbuhay (2011) reported that cosmic rays with high rigidity reached record setting intensities during this time (see also Heber *et al.,* 2009; McDonald, Webber and Reames, 2010). It followed from observations for this period that the delay between the time for minimum sunspot numbers and maximum cosmic ray intensities was at least three times longer than during previous even numbered solar cycles (see also *e.g,*. Kane, 2011; Aslam and Badruddin, 2012; Mewaldt, 2012).

Apart from this well-known 11-year cycle in the long-term records of neutron monitors (NMs), 22-year cycles also have been observed, leading to galactic CR intensity profiles being peaked at solar minimum during magnetic field polarity cycles known as $A < 0$ (around 1965, 1987 and again around 2009) while exhibiting flatter profiles during $A > 0$ cycles (around 1976-77 and 1997-8). It is generally accepted that the explanation for this 22-year cycle lies in the occurrence of drifts of these cosmic particles caused by the gradients and curvatures of the large-scale HMF together with the presence of a wavy HCS. This was originally illustrated with a numerical modulation model by Jokipii and Kopriva (1979). Potgieter and Moraal (1985) confirmed this prediction and pointed to a subtlety which became known as the 'cross-over' of solar minimum proton spectra (Reinecke and Potgieter, 1994 and references therein). What this means is that while galactic CR intensities at NM energies had higher levels in $A < 0$ cycles than in $A > 0$ cycles, the opposite happened at low energies at Earth (see also Webber, Potgieter and Burger, 1990). The reason is that an increased flux of positively charged low-energy CRs (mainly protons) arrive at Earth during $A > 0$ cycles when they drift inwards primarily through the polar regions of the heliosphere. Apart from gradient and curvature drifts, cosmic rays also drift inwards or outwards along the HCS depending on the polarity cycle and the charge of the particles. During $A < 0$ cycles, such as the present cycle, positively charged particles drift in towards Earth mostly along this HCS, so that its waviness as represented by the tilt angle (Hoeksema, 1991) plays a more significant modulation role than during $A > 0$ cycles (Webber, Potgieter and Burger, 1990; Potgieter and le Roux, 1992; Potgieter and Ferreira, 2001). It is therefore to be expected that proton spectra for $A < 0$ cycles should be lower than spectra during $A > 0$ cycles at energies $E < \sim 1$ GeV (see examples by Langner, Potgieter and Webber, 2004; Ngobeni and Potgieter, 2008).

The conceptual paradigm of particle drifts (read gradient and curvature drifts) dominating CR modulation especially during periods of solar minimum conditions, and in a different manner in consecutive solar cycles, has become widely accepted and popular. Of course, particle drifts is one of four major and equally important modulation mechanisms in the heliosphere; see reviews by Heber and Potgieter (2006, 2008) and Potgieter (2011). Recently, Cliver, Richardson and Ling (2011) challenged this concept, arguing that particle drifts was not at all 'dominant' during the recent $A < 0$ minimum modulation period, instead the behaviour of CRs seemed to be determined primarily by diffusion (and convection, of course). They argued that the record high galactic CR intensities observed in 2009 resulted from a reduction in the global magnitude of the HMF (from ~5.4 nT to ~3.9 nT) in comparison with the $A < 0$ minimum in 1986 rather than from mainly a reduction in the HCS tilt angle. They consequently prefer the cosmic ray magnetic field (CR-B) relation pioneered by Burlaga et al. (1985) as the main cause of global modulation, a process that is principally diffusion 'dominated'.

It is thus clear that the last period of declining and minimum solar activity, especially from 2006 to the end of 2009, and the subsequent increase in CRs deserves special attention from a theoretical and modulation modelling point of view. The purpose of this paper is to use a full three-dimensional (3D) numerical galactic CR modulation model to establish if one of the major modulation mechanisms actually 'dominated' during the past solar minimum period or whether it was an interplay of all major mechanisms. This study is made possible through the availability of comprehensive numerical models and the precise proton spectra observed by the *PAMELA* space mission as summarized below.

In this context, we present observed galactic proton spectra, from 100 MeV to 50 GeV, for the years 2006, 2007, 2008 and 2009 from the *PAMELA* space detector together with the solutions of a comprehensive 3D model for the modulation of CRs in the heliosphere. The latter is used to determine what was responsible for the modulation of protons from 2006 to solar minimum modulation in 2009, and why the observed proton spectrum for 2009, in contrast to what was discussed above, was the highest spectrum ever recorded at Earth since the beginning of the space age.

## 2. Observations from 2006 to 2009

*PAMELA* (*Payload for Antimatter Matter Exploration and Light-nuclei Astrophysics*) is a satellite-borne experiment designed for cosmic-ray antimatter studies. The instrument is flying on board the Russian *Resurs-DK1* satellite since June 2006, following a semi-polar near-Earth orbit. The experiment has provided important results on the antiproton (Adriani *et al.,* 2009a) and positron galactic abundances (Adriani *et al.,* 2009b; Boezio *et al.,* 2009). The high-resolution *PAMELA* spectrometer also allows to perform hydrogen and helium spectral measurements up to 1.2 TV (Adriani *et al.,* 2011), which is the highest limit achieved by this kind of experiments. Adriani *et al.* (2013) presented observations down to 400 MV of the absolute flux of protons from July 2006 until the end of 2009 which is important to solar modulation studies and which are used in this work. Large proton statistics collected by the instrument allowed the measurement of the proton flux for each Carrington rotation. In Figure 1 these spectra are shown from July 2006 to the very beginning of 2010. The spectrum at the end of December 2009 was the highest recorded. In January 2010, solar activity picked up significantly so that the proton intensity had begun to decrease. In order to emphasize how decreasing solar modulation conditions allowed galactic protons to increase at Earth, especially at low kinetic energies, the spectra in Figure 1 are used to calculate and plot the intensity ratios as a function of kinetic energy with respect to July 2006. This is shown in Figure 2. Note that these spectra got increasingly softer from 2006 to 2009.

Four selected proton spectra are shown in Figure 3 as observed from 100 MeV to 50 GeV by the *PAMELA* detector. The periods are: Nov. 2006, Dec. 2007, Dec. 2008 and Dec. 2009. All these spectra were thus observed during the present $A < 0$ magnetic polarity cycle. This figure further illustrates how the proton spectra unfolded from late in 2006 to the end of 2009. These four periods were selected to be studied and to be reproduced with the modulation model discussed below. For this purpose the relevant modulation conditions, as a prerequisite for applying the modulation model, have been studied and are given below.

In Figure 4 the heliospheric modulation conditions based on a selection of observations are illustrated from the year 2000 to solar minimum activity conditions before and in 2009 up to 2012. The top panel shows the sunspot number (http://sidc.oma.be/) that evidently became quite low already in 2007 to remain very low until it has begun to increase in 2009 to start a new sunspot cycle. The second panel gives the count rate of the Hermanus NM, normalized to 100% in March 1987 (http://www.nwu.ac.za/content/neutron-monitor-data) as an indication how cosmic rays at relative high rigidity responded to the decrease in solar activity. This NM data show an increase from 2006 to 2007 with a plateau in 2008 and clearly a maximum from early until late in 2009. In January 2010, also these high rigidity CRs have begun to decrease significantly and are still doing so. The third panel shows the HCS tilt angle for the same period (radial model; http://wso.stanford.edu/) which is from a modulation modelling point of view a good proxy for solar activity. Qualitatively, the message is the same but the tilt angle did not decrease as quickly as in previous solar minimum periods, remaining at relatively high values from 2006 to 2008. Only in 2009 did the tilt angle drop to values less than 10°. As happened before just after solar minimum modulation, the tilt angle has begun to increase sharply from early in January 2010, reaching 60° already in 2011. In the fourth panel the monthly (green) and daily (light green) HMF averages observed close to Earth by ACE (http://nssdc.gsfc.nasa.gov/) are shown as a function of time. In contrast to previous solar minima, the averaged HMF value dropped to ~4 nT for an extended period of time and sporadically even below this value.

The changes in the tilt angle and HMF values are relevant for modelling and are emphasized in Figure 5 by plotting them on a larger scale for the period 2006 to 2010. In order to produce a proton spectrum with the 3D model for a specific period such as Nov. 2006, the tilt angle and HMF averaged values are required for the preceding months as input parameters to the model. Because galactic CRs respond to modulation conditions between the modulation boundary and the point of observation, the average tilt angle between Earth and the boundary must be estimated based on what had been observed at Earth about a year before the observational time of the particular proton spectrum. We therefore calculate the average tilt angle and HMF values as shown in Figure 5, and listed in the first two rows of Table 1 for 2006, 2007, 2008 and 2009, respectively. These averaged values are used as observationally motivated input to our model, with the diffusion coefficients and drift coefficient (being theoretically motivated) recalculated every time for the four selected time periods. The period over which the averaging is done has to be long enough to assure that these modulation conditions are settled throughout the heliosphere. The coefficients and particularly their rigidity dependence are discussed in section 4.

## 3. Numerical model and modulation parameters

A full three-dimensional (3D) model is used to compute the differential intensity of 10 MeV to 30 GeV cosmic ray protons at Earth. It is based on the numerical solution of the well-known heliospheric transport equation (TPE; Parker, 1965)

$$\frac{\partial f}{\partial t} = -\left(\mathbf{V}+\langle\mathbf{v}_D\rangle\right)\cdot\nabla f + \nabla\cdot\left(\mathbf{K}_s\cdot\nabla f\right) + \frac{1}{3}(\nabla\cdot\mathbf{V})\frac{\partial f}{\partial \ln P}, \tag{1}$$

where $f(\mathbf{r},P,t)$ is the cosmic ray distribution function, $P$ is rigidity, $t$ is time, $\mathbf{r}$ is the position in 3D, with the usual three coordinates $r$, $\theta$, and $\varphi$ specified in a heliocentric spherical coordinate system where the equatorial plane is at a polar angle of $\theta = 90°$. It is assumed that $\partial f / \partial t = 0$, which means all short-term modulation effects (periods shorter than one solar rotation) are neglected, a most reasonable assumption for solar minimum conditions. Terms on the right hand side respectively represent convection, with $\mathbf{V}$ the solar wind velocity; averaged particle drift velocity $\langle\mathbf{v}_D\rangle$ caused by gradients and curvatures in the global HMF; diffusion, with $\mathbf{K}_s$ the symmetric diffusion tensor and then adiabatic energy changes, with $(\nabla\cdot\mathbf{V}) > 0$ giving adiabatic energy losses except inside the heliosheath where it is assumed that $(\nabla\cdot\mathbf{V}) = 0$ (see also *e.g.*, Langner *et al.*, 2006; Nkosi, Potgieter and Webber, 2011). Adiabatic energy change is one of the four major modulation mechanisms and is crucially important for galactic CR modulation in the inner heliosphere (see the illustrations by Strauss, Potgieter and Ferreira, 2011; Strauss *et al.*, 2011).

For clarity on the role of diffusion, particle drifts, convection, and adiabatic energy loss, the TPE can be written in terms of a heliocentric spherical coordinate system as follows:

$$\frac{\partial f}{\partial t} = \overbrace{\left[\frac{1}{r^2}\frac{\partial}{\partial r}\left(r^2 K_{rr}\right) + \frac{1}{r\sin\theta}\frac{\partial K_{\phi r}}{\partial \phi}\right]}^{diffusion}\frac{\partial f}{\partial r} + \overbrace{\left[\frac{1}{r^2\sin\theta}\frac{\partial}{\partial\theta}\left(K_{\theta\theta}\sin\theta\right)\right]}^{}\frac{\partial f}{\partial \theta}$$

$$+ \overbrace{\left[\frac{1}{r^2\sin\theta}\frac{\partial}{\partial r}\left(rK_{r\phi}\right) + \frac{1}{r^2\sin^2\theta}\frac{\partial K_{\phi\phi}}{\partial \phi} + \Omega\right]}^{diffusion}\frac{\partial f}{\partial \phi}$$

$$+ \overbrace{\left[K_{rr}\frac{\partial^2 f}{\partial r^2} + \frac{K_{\theta\theta}}{r^2}\frac{\partial^2 f}{\partial \theta^2} + \frac{K_{\phi\phi}}{r^2\sin^2\theta}\frac{\partial^2 f}{\partial \phi^2} + \frac{2K_{r\phi}}{r\sin\theta}\frac{\partial^2 f}{\partial r\partial \phi}\right]}^{diffusion}$$

$$+\left[-\langle v_d\rangle_r\right]\frac{\partial f}{\partial r}+\left[-\frac{1}{r}\langle v_d\rangle_\theta\right]\frac{\partial f}{\partial \theta}+\left[-\frac{1}{r\sin\theta}\langle v_d\rangle_\phi\right]\frac{\partial f}{\partial \phi}\overbrace{\phantom{X}}^{drift}$$

$$\overbrace{-V\frac{\partial f}{\partial r}}^{convection}+\overbrace{\frac{1}{3r^2}\frac{\partial(r^2 V)}{\partial r}\frac{\partial f}{\partial \ln P}}^{adiabatic\ energy\ changes}+\overbrace{Q_{source}}^{sources}. \quad (2)$$

Here, $K_{rr}, K_{r\theta}, K_{r\phi}, K_{\theta r}, K_{\theta\theta}, K_{\theta\phi}, K_{\phi r}, K_{\phi\theta}$ and $K_{\phi\phi}$ are the nine elements of the 3D diffusion tensor, based on a parkerian type HMF with a radial solar wind speed $V$. Note that $K_{rr}, K_{r\phi}, K_{\theta\theta}, K_{\phi r}, K_{\phi\phi}$ describe the diffusion processes and that $K_{r\theta}, K_{\theta r}, K_{\phi\theta}, K_{\theta\phi}$ describe particle drifts, in this case considered to be gradient and curvature drifts. The diffusion coefficients of special interest in a 3D heliocentric spherical coordinate system are

$$K_{rr} = K_\| \cos^2\psi + K_{\perp r}\sin^2\psi, \text{ and } K_{\theta\theta} = K_{\perp\theta}, \quad (3)$$

$$K_{\phi\phi} = K_{\perp r}\cos^2\psi + K_\|\sin^2\psi \text{ and } K_{\phi r} = \left(K_{\perp r} - K_\|\right)\sin\psi\cos\psi = K_{r\phi}. \quad (4)$$

where $K_{rr}$ is the effective radial diffusion coefficient, a combination of the parallel diffusion coefficient $K_{//}$ and the radial perpendicular diffusion coefficient $K_{\perp r}$, with $\psi$ the spiral angle of the average HMF with the radial direction; $K_{\theta\theta} = K_{\perp\theta}$ is the effective perpendicular diffusion coefficient in the polar direction. Here, $K_{\phi\phi}$ describes the effective diffusion in the azimuthal direction and $K_{\phi r}$ is the diffusion coefficient in the $\phi r$-plane, etc. Both are determined by the choices for $K_{//}$ and $K_{\perp r}$. Beyond ~20 AU in the equatorial plane $\psi \to 90°$, so that $K_{rr}$ is dominated by $K_{\perp r}$ whereas $K_{\phi\phi}$ is dominated by $K_{//}$, but only if the HMF is parkerian in its geometry.

The diffusion coefficient parallel to the averaged background HMF is given by

$$K_\| = (K_\|)_0 \beta\left(\frac{B_0}{B}\right)\left(\frac{P}{P_0}\right)^a \left(\frac{(\frac{P}{P_0})^c + (\frac{P_k}{P_0})^c}{1+(\frac{P_k}{P_0})^c}\right)^{\frac{(b-a)}{c}}, \quad (5)$$

with $\beta = v/c$, the ratio of the particle's speed to the speed of light. Here, $(K_\|)_0$ is a constant in units of $10^{22}$ cm$^2$.s$^{-1}$, with the rest of the equation written to be dimensionless with $P_0 = 1$ GV and $B$ the HMF magnitude with $B_0 = 1$ nT (so that the units remain $cm^2.s^{-1}$). Here $a$ is a power index that can change with time (*e.g.* from 2006 to 2009) as shown in Table 1; $b = 1.95$ and together with $a$ determine the slope of the rigidity dependence respectively above and below a rigidity with the value $P_k$, whereas $c = 3.0$ determines the smoothness of the transition. This means that the rigidity dependence of $K_{\parallel}$ is basically a combination of two power laws which will be illustrated later when the computed proton spectra are displayed. The value $P_k$ determines the rigidity where the break in the power law occurs and the value of $a$ determines the slope of the power law at rigidities below $P_k$ (see Table 1 and Figure 7 which will be discussed below).

Perpendicular diffusion in the radial direction is assumed to be given by

$$K_{\perp r} = 0.02 K_{\parallel}. \qquad (6)$$

This is a straightforward, reasonable and widely used assumption. On the other hand, the role of polar perpendicular diffusion, $K_{\perp \theta}$, in the inner heliosphere has become increasingly better understood over the past decade since it was realized that it should be anisotropic, with reasonable consensus that $K_{\perp \theta} > K_{\perp r}$ away from the equatorial regions (see also Potgieter, 2000). Advances in diffusion theory and predictions for the heliospheric diffusion coefficients (*e.g.*, Teufel and Schlickeiser, 2002; Shalchi, 2009) make it possible to narrow down the parameter space used in typical modulation models, in particular the rigidity dependence at Earth. Numerical modelling shows explicitly that in order to explain the small latitudinal gradients observed for protons by *Ulysses* during solar minimum modulation in 1994, an enhancement of latitudinal transport with respect to radial transport is required (see *e.g.,* Heber and Potgieter, 2006, and reference therein). The perpendicular diffusion coefficient in the polar direction is thus assumed to be given by

$$K_{\perp \theta} = 0.02 K_{\parallel} f_{\perp \theta} \qquad (7)$$

$$\text{with } f_{\perp \theta} = A^+ \mp A^- \tanh\left[8(\theta_A - 90^0 \pm \theta_F)\right]. \qquad (8)$$

Here, $A^{\pm} = (d \pm 1)/2$, $\theta_F = 35^0$, $\theta_A = \theta$ for $\theta \leq 90^0$ but $\theta_A = 180^0 - \theta$ with $\theta \geq 90^0$ and $d = 3.0$. This means that $K_{\theta\theta} = K_{\perp \theta}$ is enhanced towards the poles by a factor $d$ with respect to the value of $K_{\parallel}$ in the equatorial regions of the heliosphere (as also applied and motivated by Potgieter, 2000; Ferreira *et al.,* 2001, 2003; Moeketsi *et al.,* 2005; Ngobeni and Potgieter, 2011).

The pitch angle averaged guiding centre drift velocity for a near isotropic cosmic ray distribution is given by <**v**$_D$> = $\nabla \times (K_A \mathbf{e}_B)$, with $\mathbf{e}_B = \mathbf{B}/B_m$ where $B_m$ is the magnitude of the modified background HMF assumed to have a basic Parkerian geometry in the equatorial plane but modified in the polar regions (see also Potgieter, 2000). Under the assumption of weak scattering, the drift coefficient is straightforwardly given as

$$(K_A)_{ws} = (K_A)_0 \frac{\beta P}{3 B_m} \tag{9}$$

where $(K_A)_0$ is dimensionless. In this case $(K_A)_0 = 1.0$ describes what Potgieter *et al*, (1989) called 100% drifts (i.e., full 'weak scattering' gradient and curvature drifts). A deviation from this weak scattering form is given by

$$K_A = (K_A)_0 \frac{\beta P}{3B} \left( \frac{(\frac{P}{P_{A0}})^2}{1+(\frac{P}{P_{A0}})^2} \right). \tag{10}$$

This means that below $P_{A0}$ (in GV), particle drifts are progressively reduced with respect to the weak scattering case as shown in Figure 8. This is required to explain the small latitudinal gradients at low rigidities observed by *Ulysses* (Heber and Potgieter, 2006; Di Simone *et al.*, 2011). Drift velocity components in terms of $K_{r\theta}, K_{\theta r}, K_{\phi\theta}, K_{\theta\phi}$ are

$$\langle \mathbf{v}_d \rangle_r = -\frac{A}{r \sin\theta} \frac{\partial}{\partial \theta} (\sin\theta K_{\theta r}),$$

$$\langle \mathbf{v}_d \rangle_\theta = -\frac{A}{r} \left[ \frac{1}{\sin\theta} \frac{\partial}{\partial \phi} (K_{\phi\theta}) + \frac{\partial}{\partial r} (r K_{r\theta}) \right], \tag{11}$$

$$\langle \mathbf{v}_d \rangle_\phi = -\frac{A}{r} \frac{\partial}{\partial \theta} (K_{\theta\phi}),$$

with $A = \pm 1$; when this value is positive (negative), an $A > 0$ ($A < 0$) polarity cycle is described. The present polarity cycle is indicated by $A < 0$, as was also the case for the years around 1965 and 1987. During such a cycle, positively charged cosmic rays are drifting into the inner heliosphere mostly through the equatorial regions, thus having a high probability of encountering the wavy HCS directly.

The magnitude of the solar wind velocity **V** up to the heliopause, at 120 AU for this study, is given by

$$V(r,\theta) = V_0\{1-\exp[13.33(\frac{r_{sun}-r}{r_0})]\}\{1.475 \mp 0.4\tanh[6.8(\theta-\frac{\pi}{2}\pm\theta_T)]\}$$
$$[\frac{(s+1)}{2s} - \frac{(s-1)}{2s}\tanh(\frac{r-r_{TS}}{L})],$$
(12)

where $V_0 = 400$ km.s$^{-1}$, $r_{sun} = 0.005$ AU and $r_0 = 1$ AU, $s = 2.5$ and $L = 1.2$ AU. The top and bottom signs respectively correspond to the northern ($0 \leq \theta \geq \pi/2$) and southern hemisphere ($\pi/2 \leq \theta \geq \pi$) of the heliosphere, with $\theta_T = \alpha + 15\pi/180$ and $\alpha$ the tilt angle of the HCS. This determines at which polar angle the solar wind speed changes from a slow to a fast region. Here, $L$ is the shock scale length and $s$ is the compression ratio of the termination shock (TS) which changed position from 88 AU in 2006, to 86 AU for 2007, to 84 AU in 2008 and to 80 AU in 2009 in this study. The re-acceleration effects of the TS are neglected for the galactic cosmic rays considered here. However, the modulation effect of the heliosheath, which is 26 AU wide in our model, has been incorporated self consistently through the expressions for the diffusion coefficients (when $V$ drops at the TS, $B$ increases and thus changes the diffusion coefficients accordingly by the factor $s$). Beyond the TS, the solar wind profile is assumed to change so that $(\nabla \cdot \mathbf{V}) = 0$ in Equation (1). This means that no adiabatic energy changes occur for these galactic protons beyond a few AU from the TS. This might be an over simplification but it has been shown (Langner *et al.*, 2006) that the effect is insignificant for galactic CRs. For elaborated discussions on this issue, see also Strauss, Potgieter, and Ferreira (2011) and Ngobeni and Potgieter (2010). The numerical procedure used to solve Equation (1) was described by Ferreira and Potgieter (2002) and Ferreira *et al.* (2001, 2003) and Nkosi, Potgieter and Ferreira (2008). Graphical illustrations of the diffusion coefficients and the drift coefficient as a function of rigidity will be shown below, together with the computed spectra.

A galactic proton spectrum, or more specifically a local interstellar spectrum (LIS), has to be specified in modulation models as an initial condition, to be used as the input spectrum which then is modulated from a given heliospheric boundary up to Earth. The value of a 'very' LIS below a few-GeV has always been rather contentious. Obviously, the present Voyager 1 (at ~122 AU and perhaps at the heliopause, HP) and Voyager 2 (at 100AU) observations in the inner heliosheath give the lower limit to what can be called a HP spectrum which may already be close to a very LIS for protons at these energies. Unfortunately, studying proton spectra at Earth cannot assist in determining the spectral shape of the galactic or even the very LIS below ~10 GeV, because of the significant adiabatic energy losses CRs experience when traversing the heliosphere on their way to Earth (see also Strauss *et al.,* 2011b). This is a topic which deserved dedicated research starting from the distribution of galactic sources of protons, involving galactic propagation processes up to what flux eventually arrives at the heliopause. We do not elaborate much further on this issue but instead present a proton LIS based on the galactic proton spectrum by Langner and Potgieter (2004) but modified at high rigidities (30-50 GV) to follow the *PAMELA* proton observations at these rigidities. This gives the differential intensity as

$$J_{LIS} = \begin{cases} 0.707 \exp(4.64 - 0.08(\ln E)^2 - 2.91\sqrt{E}) & \text{if } E < 1.4 \text{ GeV} \\ 0.685 \exp(3.22 - 2.78(\ln E) - 1.5/E) & \text{if } E \geq 1.4 \text{ GeV} \end{cases}, \quad (13)$$

in units of particles $m^{-2}.s^{-1}.sr^{-1}.MeV^{-1}$, with $E$ the kinetic energy in GeV. This LIS will be shown together with the modelled spectra in figures to follow.

Another assumption to be made in modulation models is at what rigidity does solar modulation actually starts. The *PAMELA* proton observations between 10 to 100 GeV indicate that progressive changes in the spectral shape become clearly evident below ~30 GeV. We interpret this as an indication that CRs begin to experience notable (and computable) modulation below 30 GeV although the modulation effects on spectra become significantly evident and meaningful with $E < $ ~10 GeV as illustrated in Figure 2.

Together with the question of what exactly the very LIS is at low-energies (1 MeV to 1 GeV), comes the question of where is the heliospheric modulation boundary, i.e. where exactly in spatial terms does the modulation begin? Recently, Scherer *et al.* (2011) argued that the solar modulation of galactic CRs might start beyond the HP which so far has always been considered the modulation boundary. Since these issues cannot yet be settled, we continue to simply assume the HP as the boundary where the modulation of galactic CR modulation essentially starts, assumed for this work to be at 120 AU. Only future *Voyager* 1 and *Voyager* 2 observations will reveal what the situation is and what the very LIS proton spectrum may be.

## 4. Modelling results

In order to solve the TPE as applicable to Nov. 2006, Dec. 2007, Dec. 2008 and Oct.-Dec. 2009, the calculated tilt angles and observed HMF magnitude at Earth were averaged as discussed above and shown in Figure 5. In our modelling approach, this produces the averaged modulation conditions in the heliosphere up to about a year before the proton spectrum for a given time was observed at Earth.

In Figure 6 the *PAMELA* proton spectra, as observed from 2006 (blue symbols) to 2009 (red symbols), are overlaid by the corresponding computed spectra (solid lines). During this time the HCS tilt angle changed from an averaged value of $\alpha$ ~15.7° to ~10.0°, with an accompanying change in the averaged HMF at Earth from $B$ ~5.05 nT to ~3.94 n T. This illustrates that the model can reproduce the observed proton spectra for the selected periods in 2006 to 2009 when the modulating effects of the various parameters as described above are combined. Changes to the modulation parameters required to do this are summarized in Table 1, in particular the changes to the rigidity dependence of the diffusion coefficients with time, also illustrated in Figure 7. The corresponding small changes in the rigidity dependence of the drift coefficient are shown in Figure 8.

Figure 9 highlights how CR proton spectra had increased with respect to Nov. 2006. The ratio of the differential intensities relative to Nov. 2006 is depicted for the selected period in 2007, 2008 and 2009, comparing the observed ratios with the corresponding modelled ratios. It follows that at 1 GeV, the proton differential intensity increased over this period by a factor of ~1.5 whereas at 100 MeV it increased by a factor of ~3.

The extent to which diffusion and particle drifts contributed to the total observed modulation from 2006 to 2009 was also investigated, the results of which are given in Figure 10 as differential intensities as a function of time. The results in this figure can be obtained only with a numerical model in which individual processes can be switched on and off for illustrative purposes. It shows the computed relative contributions of diffusion and particle drifts in comparison to the observed modulation of CR protons from the end of 2006 to the end of 2009. The details are as follows. The bottom curve shows how the model responds to changing only the averaged HCS tilt angle, as listed in Table 1, while every other modulation parameter, in particular the HMF magnitude, was kept unchanged. The second curve from the bottom (dashed line) shows how the model responds to changing only the diffusion coefficients, that is, with the tilt angle and the drift coefficient unchanged. These changes are caused by the decreasing averaged HMF magnitude from 2006 to 2009 as listed in Table 1.

The third curve from the bottom (dotted-dashed line) is how the model responds when both the tilt angle and diffusion coefficients are changed but nothing else. In effect, this is the combination of the first two curves. The top curve illustrates how the model responds when global and gradient drifts are also changed by varying $B$ in Equation (10), together with all previous changes. Evidently, by doing so, the observed proton intensities were matched and reproduced by the model.

## 5. Discussion

The aim of the study was to reproduce a selection of four monthly-averaged proton spectra, between Nov. 2006 and Dec. 2009, each of which is separated by approximately 1 year as observed by the *PAMELA* space experiment. This was done by using ~1 year antecedent averaged values for $\alpha$ and $B$ in the numerical model, respectively 15.7°, 14.0°, 14.3° and 10.0°; and 5.05 nT, 4.50 nT, 4.25 nT and 3.94 nT. In order to accurately represent the four observed spectra, as shown in Figure 6, the value of the diffusion coefficients at Earth had to be increased, but more significantly, their rigidity dependence had to be adjusted (to be less rigidity dependent at lower energies) as illustrated in Figure 7 for every year because the spectra became progressively softer as shown in Figure 3. The rigidity dependence of the drift coefficient had to be changed very little as shown in Figure 8. This means that even under perfect solar minimum conditions, the weak scattering approach to drifts at low energies is too simple and that the modification as given by Equation (10) is justified. Additional global drifts at these low energies were thus not required to allow for the softening of the 2009 spectrum.

Adjusting the ratio between the parallel and the two perpendicular diffusion coefficients was also not necessary but the enhancement of the polar dependence of perpendicular diffusion in the polar direction was still required in order to produce the very small latitudinal gradients reported by De Simone *et al.* (2011) when comparing *PAMELA* observations to that of *Ulysses*.

The changes in the rigidity dependence of the proton mean free paths (MFPs) can be summarized as follows. At rigidities below ~2 GV, the MFPs changed from a $P^{0.56}$ dependence in 2006 to a $P^{0.48}$ in 2007, and from a $P^{0.39}$ dependence in 2008 to a $P^{0.28}$ in 2009. At 1 GV, for example, $\lambda_\parallel$ increased by a factor of ~2.3, from ~0.13 AU in 2006, to ~0.3 AU in 2009. Above ~5 GV, the MFPs had a steady $P^{1.95}$ dependence during this time. It follows from these results that the CR proton intensities increase substantially at lower energies ($E <$

~1 GeV), as shown in Figure 9, which, in order to reproduce the *PAMELA* spectra, requires corresponding decreases in the rigidity dependence of $\lambda_{\parallel}$, $\lambda_{\perp r}$ and $\lambda_{\perp \theta}$.

The assessment made by Cliver*, Richardson and Ling (2011)* that the 2009 modulation minimum could be described as more 'diffusion dominated' than previous solar minima seems correct but not in the sense that particle drifts did not play a role as they concluded. In fact, as we illustrate, it had played a significant role but the observable modulation effects were not as well correlated with the waviness of the HCS as before. However, the protons still experienced significant and increasing global gradient and curvature drifts as the HMF decreased from 2006 to the end of 2009.

From Figure 10 follows that at the end of 2009, the decrease in the tilt angle (HCS drifts) only accounted for 25% of the total intensity increase at 1 GeV from 2006 to 2009 (factor increase of 1.10) while changes only in the diffusion coefficients (no change in drifts) accounted for 50% (factor increase of 1.22). When the two effects are combined, it accounts for ~68% (factor increase of 1.30). To take this to the 100% level (factor increase of 1.44), global drifts had to be changed according to changes in the value of *B*. This can roughly be summarized that diffusion contributed ~50% of the total cosmic proton intensities observed at Earth while particle drifts contributed the other 50%. However, this study serves to illustrate the importance of each of the modulation processes, even adiabatic cooling that determines the slope of the spectra below ~500 MeV.

Based on our modelling results, it has become evident that if the same quiet modulation conditions would again occur during the next solar minimum, an $A > 0$ polarity cycle, it is expected that a higher proton spectrum than in December 2009 will be observed at kinetic energies less than a few GeV.

6. Summary and conclusions

Since the beginning of the space age, the highest cosmic ray proton spectrum has been observed by *PAMELA* in December 2009. This was unexpected because during previous $A < 0$ polarity cycles, proton spectra were always lower than for $A > 0$ cycles at kinetic energies less than a few-GeV, in full accord with drift models.

Using *PAMELA* proton observations between 10 GeV and 100 GeV an adjusted local interstellar spectrum was constructed, also indicating that CRs should be considered modulated from 30 GeV and lower.

From November 2006 to December 2009, the proton spectra (shown from 100 MeV to 50 GeV) became significantly softer with increasingly more low energy protons reaching Earth. In order to simulate this observation, the rigidity dependence of the diffusion coefficients had to decrease significantly.

The 2009 modulation minimum period can be described as more 'diffusion dominated' than previous solar minima but not in the sense that particle drifts did not play a role. We illustrate that drifts had played a significant role but that the observable modulation effects were not as well correlated with the waviness of the HCS as during previous $A < 0$ polarity cycles. However, the CR protons still experienced significant and increasing gradient and curvature drifts as the HMF decreased significantly from 2006 to the end of 2009. Generally, we find that diffusion contributed ~50% of the total cosmic proton intensities observed at Earth while particle drift contributed the other ~50%. This study serves to illustrate the importance of each of the modulation processes. It is thus concluded that all modulation processes contributed to the observed changes in the proton energy spectra.

The next phase of this study is to apply the model to the electron spectra observed by *PAMELA* for the same periods. If electron spectra at the same rigidity as protons could also be obtained, a study can be performed about how the electron to proton ratio changed from 2006 to 2009. If positrons could also become available, such a study should contribute further and even clarify the role that drifts played during this unusual solar minimum. We conclude that the 2009 solar minimum modulation period was unusual and clearly different than previous A < 0 polarity minima.

**Acknowledgements** The South African authors express their gratitude for the partial funding granted by the South African National Research Foundation (NRF) under the Italy-South Africa Research Cooperation Programme. They also thank the INFN for the partial financial support when visiting in Trieste and Rome over a span of three years.

**Figure captions**

**Figure 1:** Proton spectra, averaged over one Carrington rotation, as observed by the *PAMELA* space instrument from July 2006 to the beginning of 2010 (see the colour coding on the right). The spectrum at the end of December 2009 was the highest recorded.

**Figure 2:** Proton intensity ratios as a function of kinetic energy and with respect to July 2006 (see the colour coding on the right) calculated from the spectra shown in Figure 1. The spectrum at the end of December 2009 was the highest recorded. It is assumed that no solar modulation occurs above 50 GeV.

**Figure 3:** Galactic proton spectra observed by *PAMELA* for four selected periods: Nov. 2006, Dec. 2007, Dec. 2008 and Dec. 2009. These spectra were observed in the present $A < 0$ magnetic polarity cycle.

**Figure 4:** An illustration of heliospheric modulation conditions from the year 2000 (indicated by 00) to solar minimum activity in 2009, and up to 2012. Top panel shows sunspot number (http://sidc.oma.be/). Second panel gives the count rate of the Hermanus neutron monitor, normalized to 100% in March 1987 (http://www.nwu.ac.za/content/neutron-monitor-data). Third panel shows the HCS tilt angle (radial model; http://wso.stanford.edu/). Fourth panel gives the monthly (green) and daily (light green) HMF averages observed at Earth by ACE (http://nssdc.gsfc.nasa.gov/).

**Figure 5:** Yearly averaged values are calculated for the HCS tilt angle (top panel; blue curve) and HMF magnitude (bottom panel, green curve) as representation of estimated modulation conditions in the heliosphere for the prior year. Grey regions correspond to time frames over which averages were calculated; darker colour bands give the calculated averages, along with some margin of error.

**Figure 6:** *PAMELA* proton spectra of 2006 (blue symbols) to 2009 (red symbols), overlaid by the corresponding computed spectra (solid lines). During this time the tilt angle changed from $\alpha \sim 15.7°$ to $\sim 10.0°$, with an accompanying change in the HMF at Earth from $B$ at $\sim 5.05$ nT to $\sim 3.94$ nT.

**Figure 7:** Rigidity dependence for the parallel ($\lambda_\parallel$) and radial perpendicular mean free path ($\lambda_{\perp r}$) of protons at Earth as they changed from 2006 to 2009 in order to obtain the computed spectra shown in Figure 6. Here, $\lambda_\perp$ represents both $\lambda_{\perp r}$ and $\lambda_{\perp \theta}$. See also Equations (5), (6) and (7), and Table 1.

**Figure 8:** Particle drift scale in AU at Earth, as a function of rigidity for the selected periods in 2006 to 2009. Vertical dashed line indicates $P_{A0}$ as in Equation (10). Grey dotted line gives the weak scattering drift scale as in Equation (9).

**Figure 9:** Computed ratios of differential intensities for selected periods in 2007, 2008, 2009 with respect to Nov. 2006 as a function of kinetic energy in comparison to what was observed.

**Figure 10:** Computed differential intensity for 1.0 GeV protons as a function of time, from November 2006 to December 2009. Only four values from the computed spectra in Figure 3 are plotted (black dots), along with the monthly-averaged *PAMELA* observations. Dotted and dashed lines respectively correspond to intensity levels caused by changes in only the current sheet tilt angle and only the diffusion coefficients. Dotted-dashed line represents the combined effect from these processes, where global particle drifts due to changes in the HMF magnitude (Equation 10), are excluded. Solid line represents the total computed intensity increase from the modulation processes combined.

**Table 1**

Summary of input and modulation parameters used to compute proton spectra for 2006 to 2009 that agree with the *PAMELA* observations. See Equations (6), (7), (8) and (11). Only values that changed over this period are indicated.

| Parameter | 2006 | 2007 | 2008 | 2009 |
|---|---|---|---|---|
| $\alpha$ (degrees) | 15.7 | 14.0 | 14.3 | 10.0 |
| $B$ (nT) | 5.05 | 4.50 | 4.25 | 3.94 |
| $\lambda_{\parallel}$ (AU) at Earth at 100 MV | 0.04 | 0.06 | 0.09 | 0.12 |
| $a$ | 0.56 | 0.48 | 0.39 | 0.28 |
| $P_k$ (GV) | 4.0 | 4.0 | 4.0 | 4.2 |
| $P_{A0}$ | $1/\sqrt{10}$ | $1/\sqrt{10}$ | $1/\sqrt{10}$ | $1/\sqrt{40}$ |

**FIGURES with CAPTIONS**

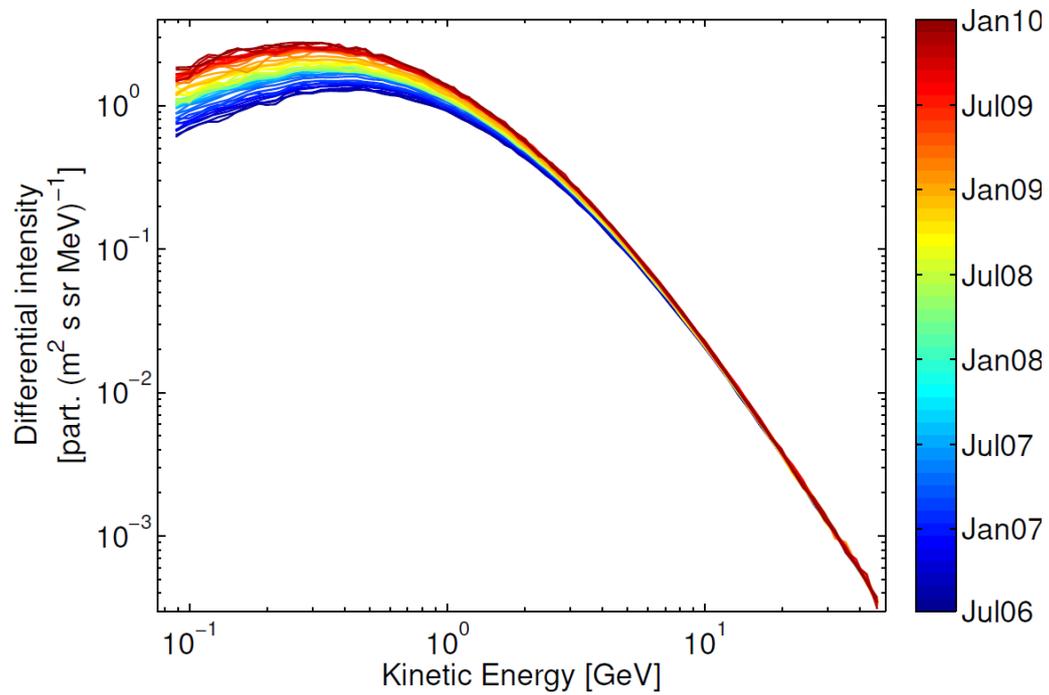

**Figure 1:** Proton spectra, averaged over one Carrington rotation, as observed by the *PAMELA* space instrument from July 2006 to the beginning of 2010 (see the colour coding on the right). The spectrum at the end of December 2009 was the highest recorded.

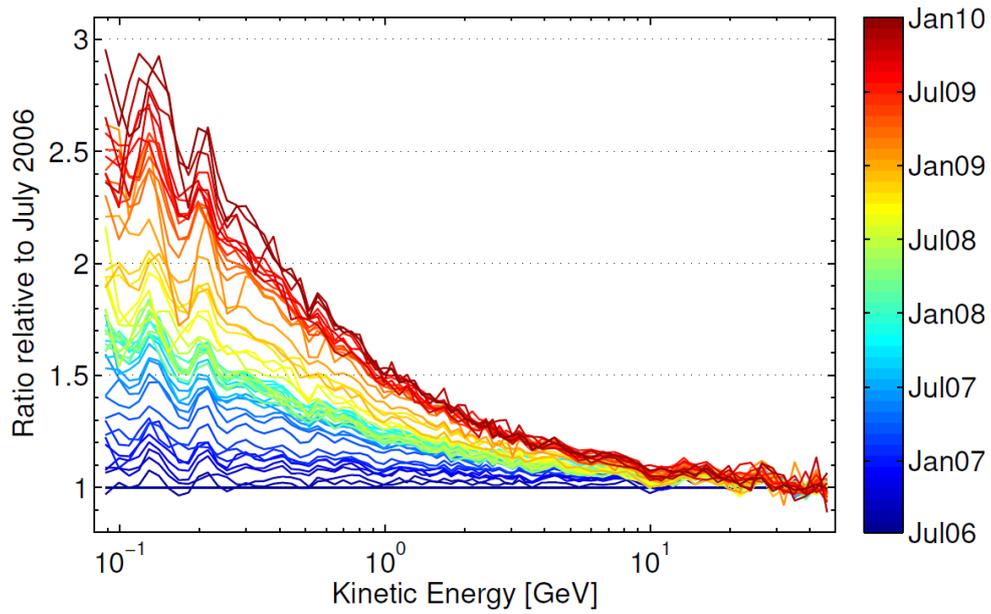

**Figure 2:** Proton intensity ratios as a function of kinetic energy and with respect to July 2006 (see the colour coding on the right) calculated from the spectra shown in Figure 1. The spectrum at the end of December 2009 was the highest recorded. It is assumed that no solar modulation occurs above 50 GeV.

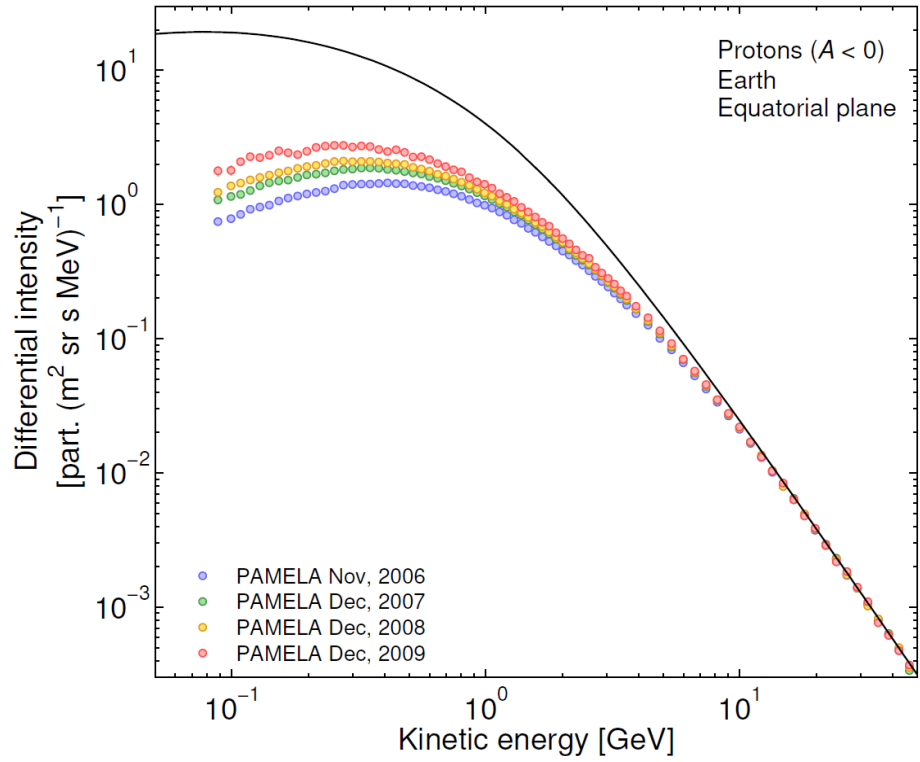

**Figure 3:** Galactic proton spectra observed by *PAMELA* for four selected periods: Nov. 2006, Dec. 2007, Dec. 2008 and Dec. 2009. These spectra were observed in the present $A < 0$ magnetic polarity cycle.

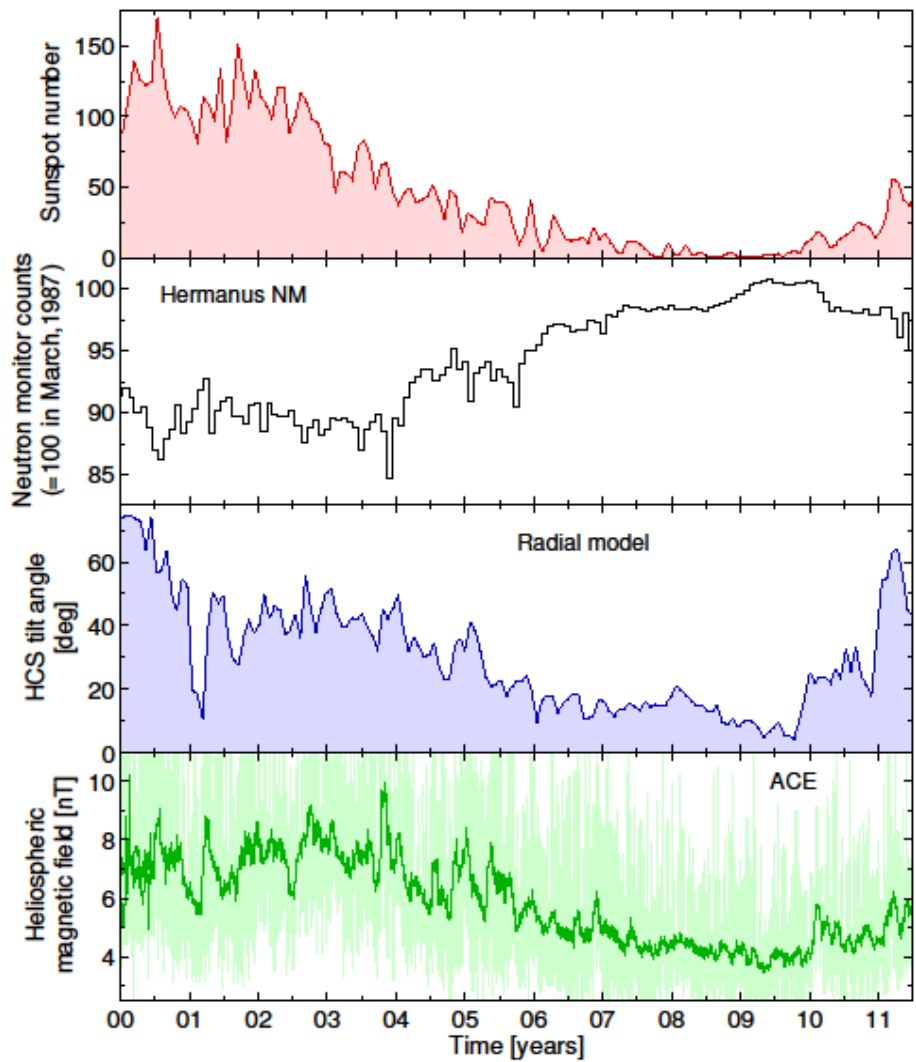

**Figure 4:** An illustration of heliospheric modulation conditions from the year 2000 (indicated by 00) to solar minimum activity in 2009, and up to 2012. Top panel shows sunspot number (http://sidc.oma.be/). Second panel gives the count rate of the Hermanus neutron monitor in South Africa at a cut-off rigidity of 4.9 GV, normalized to 100% in March 1987. Third panel shows the HCS tilt angle (radial model; http://wso.stanford.edu/). Fourth panel gives the monthly (green) and daily (light green) HMF averages observed close to Earth by ACE (http://nssdc.gsfc.nasa.gov/).

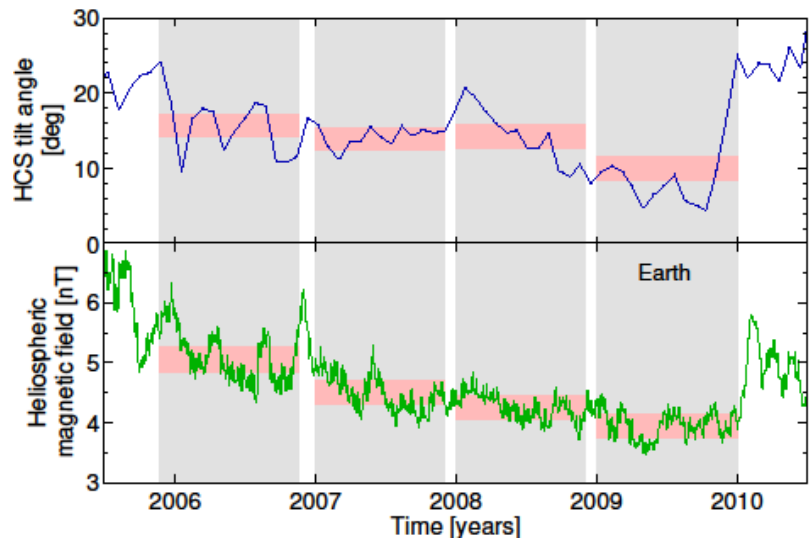

**Figure 5:** Yearly averaged values are calculated for the HCS tilt angle (top panel; blue curve) and HMF magnitude (bottom panel, green curve) as representation of estimated modulation conditions in the heliosphere for the prior year. Grey regions correspond to time frames over which averages were calculated; darker (pink) colour bands give the calculated averages, along with some margin of error.

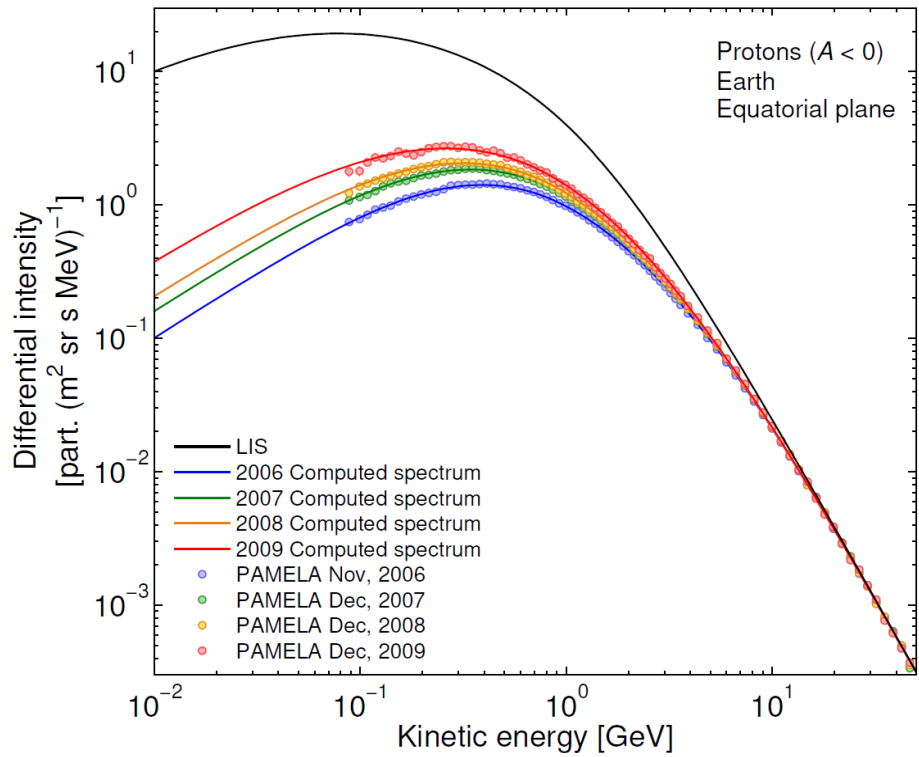

**Figure 6:** *PAMELA* proton spectra of 2006 (blue symbols) to 2009 (red symbols), overlaid by the corresponding computed spectra (solid lines). During this time the tilt angle changed from $\alpha$ ~15.7° to ~10.0°, with an accompanying change in the HMF at Earth from $B$ at ~5.05 nT to ~3.94 nT.

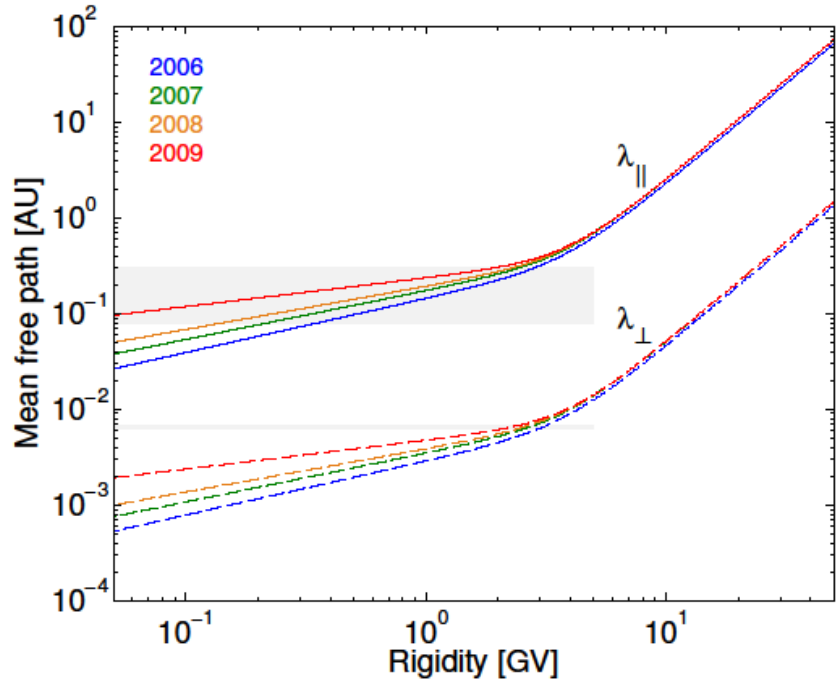

**Figure 7:** Rigidity dependence for the parallel ($\lambda_\parallel$) and radial perpendicular mean free path ($\lambda_{\perp r}$) of protons at Earth as they changed from 2006 to 2009 in order to obtain the computed spectra shown in Figure 6. Here, $\lambda_\perp$ represents both $\lambda_{\perp r}$ and $\lambda_{\perp \theta}$. See also Equations (6), (7) and (8), and Table 1.

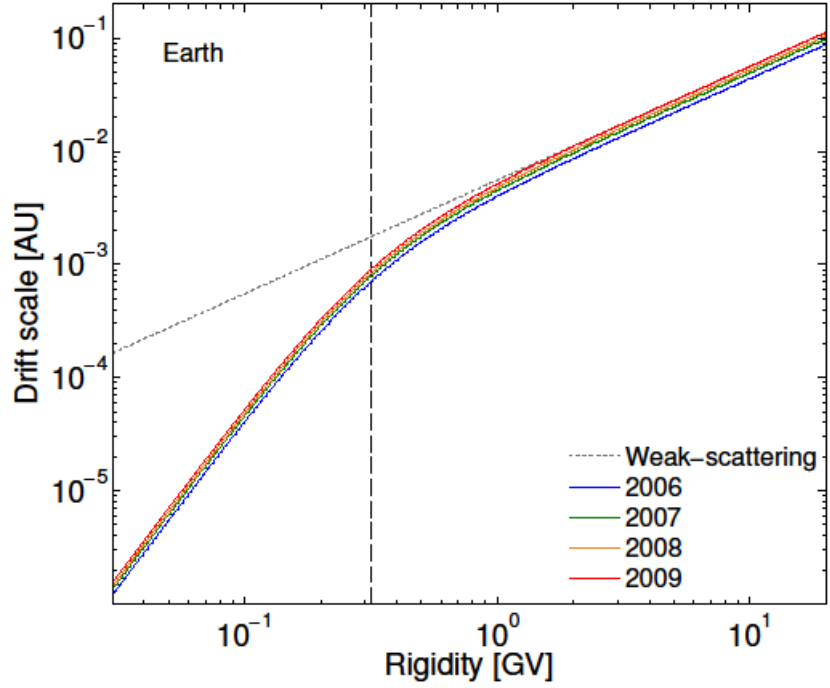

**Figure 8:** Particle drift scale in AU at Earth, as a function of rigidity for the selected periods in 2006 to 2009. Vertical dashed line indicates $P_{A0}$ as in Equation (10). Grey dotted line gives the weak scattering drift scale as in Equation (9).

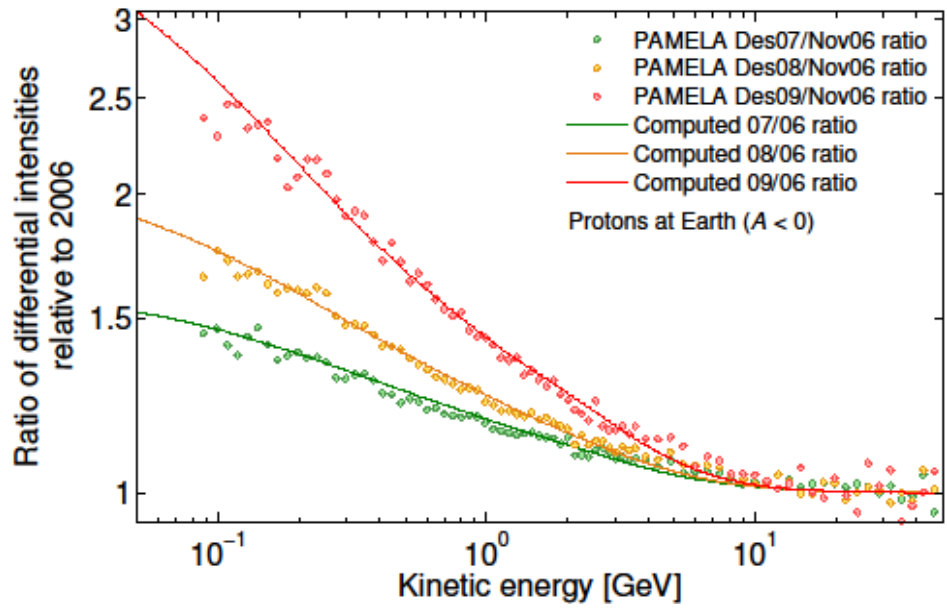

**Figure 9:** Computed ratios of differential intensities for selected periods in 2007, 2008, 2009 with respect to Nov. 2006 as a function of kinetic energy in comparison to what was observed.

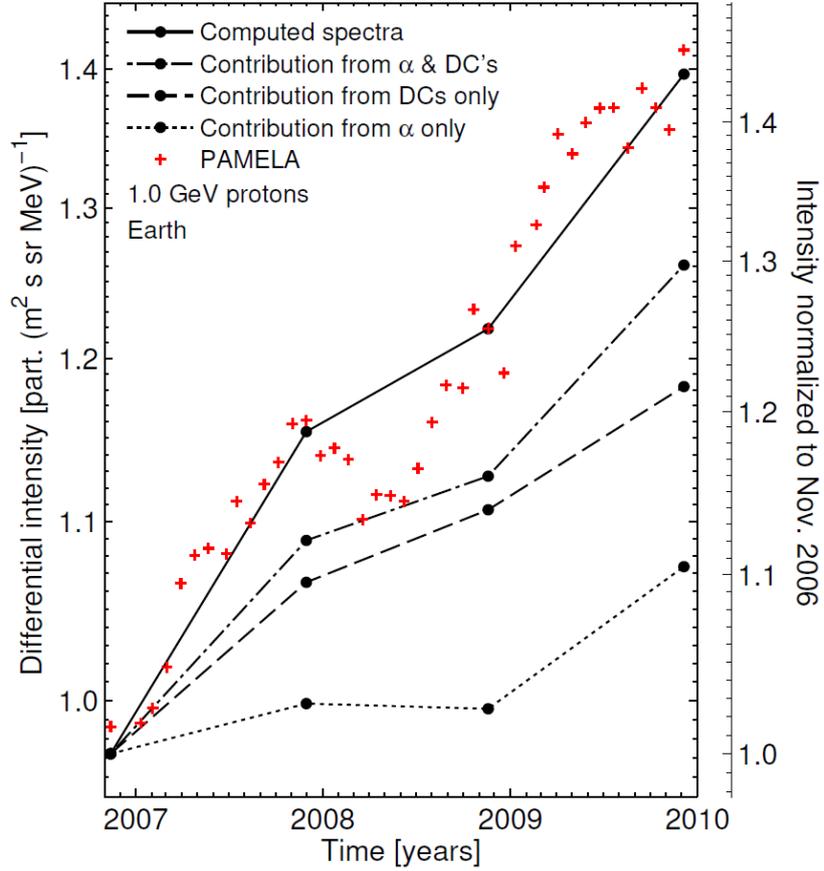

**Figure 10:** Computed differential intensity for 1.0 GeV protons as a function of time, from November 2006 to December 2009. Only four values from the computed spectra in Figure 3 are plotted (black dots), along with the monthly-averaged *PAMELA* observations. Dotted and dashed lines respectively correspond to intensity levels caused by changes in only the current sheet tilt angle and only the diffusion coefficients. Dotted-dashed line represents the combined effect from these processes, where global particle drifts due to changes in the HMF magnitude (Equation 10), are excluded. Solid line represents the total computed intensity increase from the modulation processes combined.